\documentclass[%
reprint,
amsmath,amssymb,
aps,
]{revtex4-2}

\usepackage{graphicx}
\usepackage{dcolumn}
\usepackage{bm}
\usepackage{mhchem}
\usepackage[english]{babel}

\usepackage{upgreek}

\begin{document}
	
	\preprint{APS/123-QED}
	
	\title{Spin caloritronics in a CrBr$_3$ based magnetic van der Waals heterostructure}
	
	
	\author{Tian Liu}
	\email{tian.liu@rug.nl}
	\thanks{Contributed equally to this work}
	\author{Julian Peiro}%
	\thanks{Contributed equally to this work}
	\author{Dennis K. de Wal}
	\author{Johannes C. Leutenantsmeyer}
	\author{Marcos H.D. Guimar\~aes}
	\author{Bart J. van Wees}
	\affiliation{Zernike Institute for Advanced Materials, Nijenborgh 4, 9747 AG Groningen, The Netherlands}%

	\date{\today}

	\begin{abstract}
		The recently reported magnetic ordering in insulating two-dimensional (2D) materials, such as chromium triiodide (CrI$_3$) and chromium tribromide (CrBr$_3$), opens new possibilities for the fabrication of magneto-electronic devices based on 2D systems. Inevitably, the magnetization and spin dynamics in 2D magnets are strongly linked to Joule heating. Therefore, understanding the coupling between spin, charge and heat, i.e. spin caloritronic effects, is crucial. However, spin caloritronics in 2D ferromagnets remains mostly unexplored, due to their instability in air. Here we develop a fabrication method that integrates spin-active contacts with 2D magnets through hBN encapsulation, allowing us to explore the spin caloritronic effects in these materials. The angular dependence of the thermal spin signal of the CrBr$_3$/Pt system is studied, for different conditions of magnetic field and heating current. We highlight the presence of a significant magnetic proximity effect from CrBr$_3$ on Pt revealed by an anomalous Nernst effect in Pt, and suggest the contribution of the spin Seebeck effect from CrBr$_3$. These results pave the way for future magnonic devices using air-sensitive 2D magnetic insulators.
		
	\end{abstract}

	
	\maketitle
	
	
	\section{Introduction}
	
	The search for magnetism in 2D systems has been a non-trivial topic for decades. Recently, 2D magnetism was demonstrated in an insulating material CrI$_3$ \cite{huang2017layer}, which shows antiferromagnetic exchange between the layers, resulting in zero (non-zero) net magnetization for even (odd) number of layers. It has been shown that CrBr$_3$ exhibits ferromagnetism when exfoliated down to a few layers \cite{Kim2019evolution} and monolayers \cite{Kim2019a} while preserving its magnetic order. 
	
	This discovery offers us a platform to explore magnonics in 2D systems. Magnonics refers to spintronics based on magnons, which are quantized spin waves, i.e. collective excitations of ordered electron spins in magnetic materials \cite{bloch1930theorie, Kittel2004, shan2018coupled}. Magnonic spin transport has been extensively studied in various ways in 3D magnetic insulators, e.g. spin pumping \cite{tserkovnyak2002enhanced}, Spin Seebeck Effect (SSE) \cite{uchida2010spin} and electrical injection and detection of magnons \cite{cornelissen2015long}. The outstanding magnon transport properties of the ferrimagnetic insulator yttrium iron garnet (YIG) and the robustness and fast dynamic of AFM materials like iron oxide \cite{lebrun2018tunable} and nickel oxide \cite{hahn2014conduction} triggered the development of the first magnon transport device prototypes for application using these materials \cite{cornelissen2015long,Wu2018,Chumak2014}. 
	The predicted novel physical phenomena \cite{Pershoguba2018,Cheng2016,Nakata2017,Xu2019,Ghazaryan2018} hosted by low-dimensional magnon systems represent a strong potential for 2D magnonics. Thermally excited magnon transport was reported recently in an AFM vdW 2D material MnPS$_3$ \cite{xing2019magnon}. However, magnonics in 2D van der Waals magnetic systems still remains mostly unexplored, especially for 2D ferromagnetic (FM) systems.
	
	
	One of the difficulties to study such phenomena is the easy degradation in air of the magnetic 2D materials, bringing extra technical challenges for integrating magnonic circuits with these materials. Here, we introduce a technique of bottom metallic contacts on an air-sensitive material CrBr$_3$, aiming at preliminary study of magnonics in 2D ferromagnetic materials. We select CrBr$_3$ as a medium for 2D magnonics study \cite{tsubokawa1960magnetic} as its FM order is independent on the number of layers and thus it simplifies the device fabrication. The Curie temperature of CrBr$_3$ is about T$_c$=37 $K$ \cite{tsubokawa1960magnetic} in bulk, reducing to 27 K for monolayers \cite{Kim2019a}. CrBr$_3$ presents Perpendicular Magnetic Anisotropy (PMA) \cite{Kim2019evolution} with an out-of-plane coercive field of 4 mT and an in-plane saturation field of 400 mT for a few layers \cite{Kim2019a}. The saturation magnetization of about 271 kA/m is reported nearly equal for in-plane and out-of-plane orientation in bulk and differs by less than 20\% for 3-layer CrBr$_3$ \cite{Kim2019evolution,Richter2018}.
	
	\section{Device Geometry and Measurements}
	
	In this work we employ non-local angular-dependent magnetoresistance (nlADMR) measurements on a hBN-encapsulated CrBr$_3$ flake contacted by Pt strips. ADMR measurements have been widely used to characterize the spin Hall Magnetoresistance (SMR) in local geometries  \cite{nakayama2013spin} or the spin Seebeck effect (SSE) in non-local geometry  \cite{cornelissen2015long} and identify them out of other caloritronics effects \cite{KikkawaPRB2013,Meier2015,Meyer2017,Avci2020}. We fabricated a device where Pt strips (5.5 nm thick) are deposited into a pre-etched 16.6 nm-thick hBN flake on top of a silicon oxide substrate. A 6.5 nm-thick top hBN flake is used to pick up and fully cover a 7 nm-thick CrBr$_3$ flake (about 10 layers)\cite{zomer2011transfer}. A schematic of the device and non-local measurement geometry is shown in Fig.1a. 
	
	\begin{figure*}[t]
		\includegraphics[width=440pt]{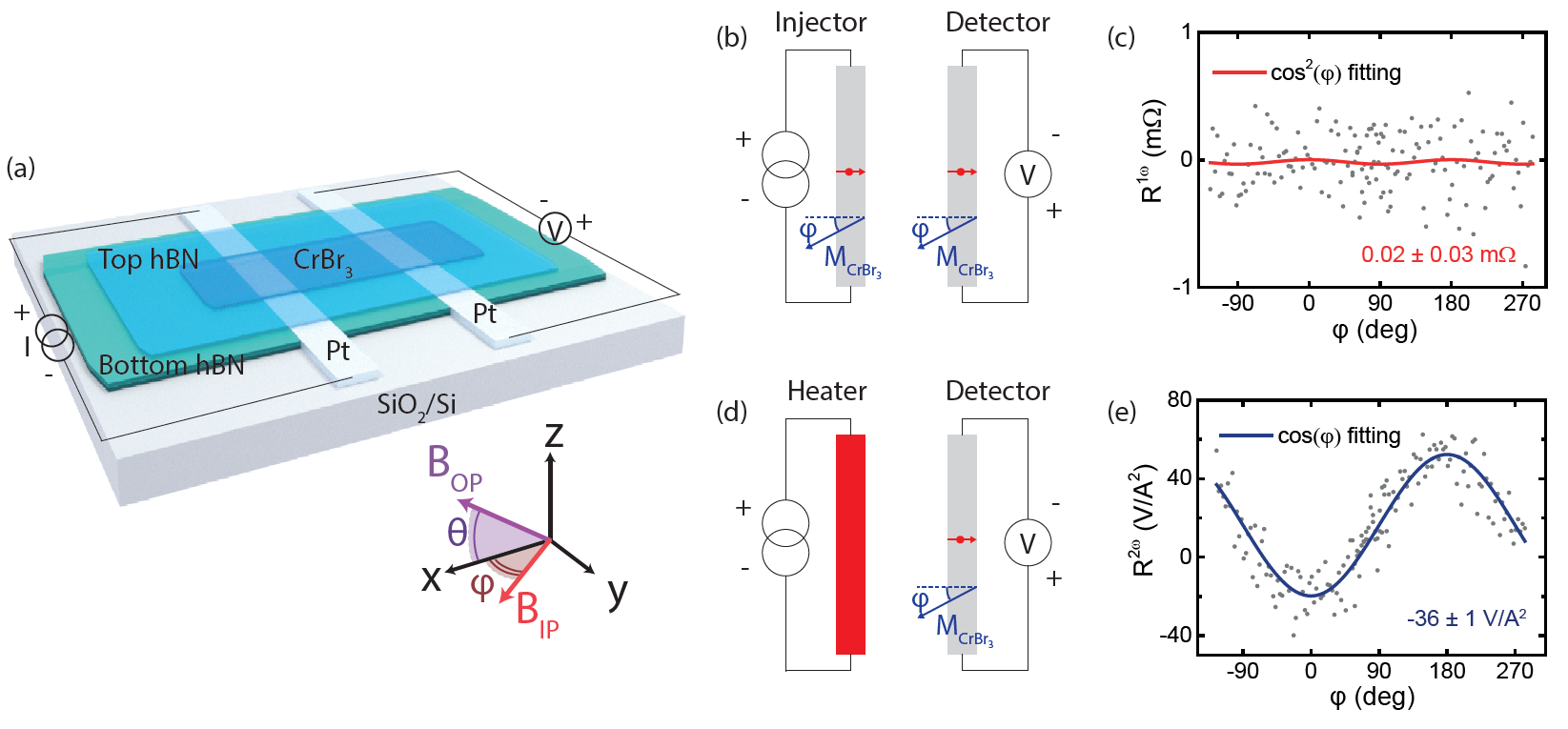}
		\caption{(a) A schematic of the device and the circuit for the non-local measurements. A 7nm-thick CrBr$_3$ flake placed on top of 5.5 nm-high Pt strips is fully encapsulated by two layers hBN. The x,y directions are defined to be in-plane (Pt strips parallel to y-axis) where the magnetic field is rotated over the azimuthal angle $\varphi$ (IP) and polar angle theta (OOP). (b) and (d) Principle of generation and detection of respectively electrically and thermally generated magnons. (c) and (e) Measured corresponding first (c) and second (e) order harmonic NL resistances with 20  $\upmu$A are fitted with the $cos^2{(\varphi)}$ and $cos{(\varphi)}$ function respectively. The small red arrows in (b,d) indicate spin polarization direction. For (e) the sign of the fitted cosine for the ISHE from the SSE agrees with this spin polarization and therefore with the standard definition of the spin Hall angle \cite{Schreier_2014}. For the measurement in Fig.\ref{fig:1}e, the offset $R_{0}^{2\omega}$ = 16.3 $\pm$ \ 0.8 \ $V/A^2$. The error bars represent the standard deviation from the fits.}
		\label{fig:1}
	\end{figure*}

	In this system, a gradient of temperature is created by the Joule heating from a remote Pt heater which generates a magnon-mediated spin flow due to the magnon density dependence on the temperature \cite{LudoThesis}, i.e. the SSE. At the interface between a magnet and a non-magnetic material, a transfer of magnon spin ($+\hbar$) from the CrBr$_3$ to the Pt is possible by spin flip of a $-\hbar/2$ spin to a $+\hbar/2$ spin in the Pt. The spin current generated this way in the Pt contact converts into a charge current by inverse spin Hall effect (ISHE) and can be measured as a voltage difference. In the geometry defined in Fig.\ref{fig:1},  the ADMR is then sensitive to the $x$ component of the magnetization of CrBr$_3$, $\mathbf{M}_x$. In the in-plane ADMR configuration (Fig.\ref{fig:1}a and d), the orientation of the magnetization with regard to the detection contact drives the angular dependence, therefore a $cos(\varphi)$ dependence is expected.
	
	%
	
	All data shown in the main text was measured on a pair formed of a 310 nm-wide injector and a 520 nm-wide detector, spaced by 500 nm edge to edge, and at a base temperature of 5K under a reference magnetic field of 4T, unless specifically mentioned. We separate different harmonics by using a standard low frequency (6Hz to 13Hz) lock-in techniques. The voltage response is composed of different orders and are expanded as: \( V\left(t\right) = R_1I\left(t\right) + R_2I\left(t\right)^2 + \cdots \) \cite{cornelissen2015long}, where \(R_i\) is the \(i\)th-order response \cite{bakker2010interplay} to the applied AC current $I\left(t\right)$. As the electrical magnon injection scales linearly with current, its response is expected in the first harmonic signal. The thermal injection depends quadratically on the applied current and the response appears in the second harmonic signal.

	\begin{figure}
		\includegraphics[width=250pt]{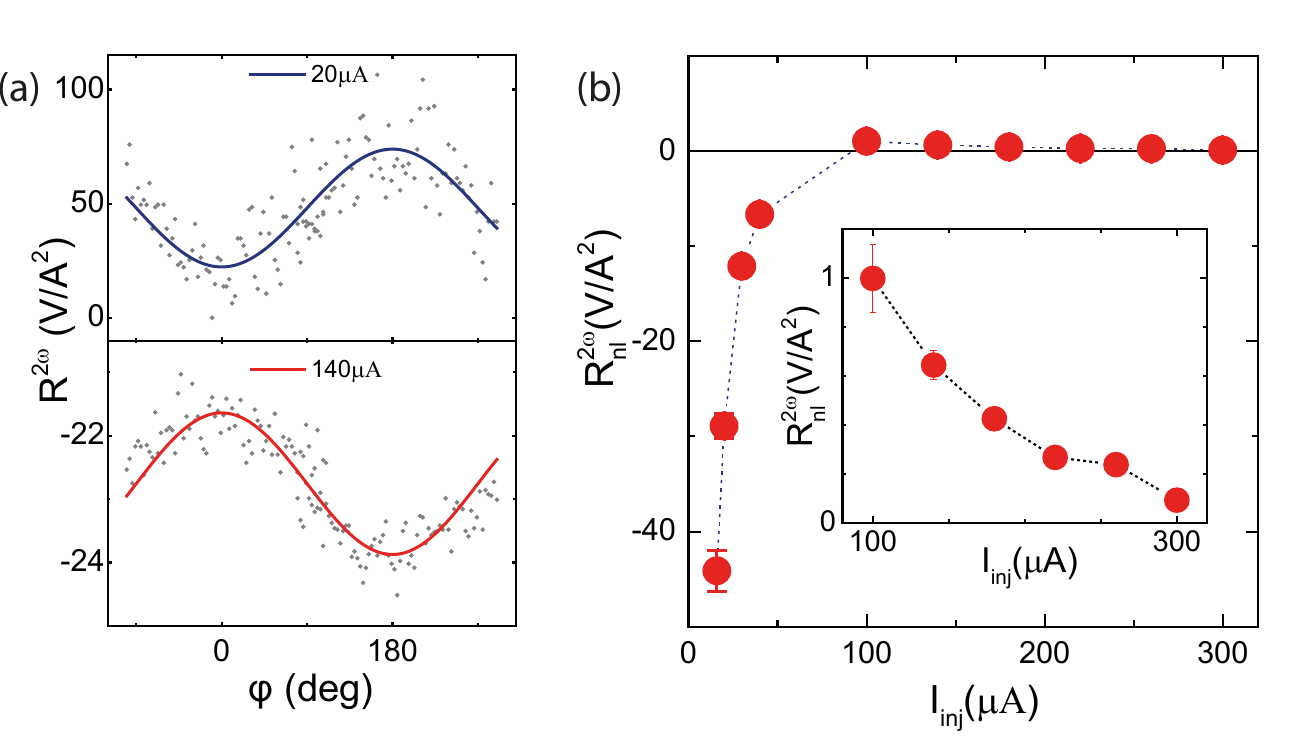}
		\caption{The dependence of second harmonic signals on applied current through the injector. (a) Top panel: low bias signals with cos($\varphi$) fitting measured at 20 $\upmu$A, with a fitted amplitude (-29\ $\pm$\ 1\  V/A$^2$); bottom panel: high bias signals with cos($\varphi$) fitting measured at 140 $\upmu$A, with a fitted amplitude (0.64\ $\pm$\ 0.03\  V/A$^2$). (b) Bias dependence of $R_{\textrm{nl}}^{2\omega}$. Bias dependence shown in these graphs were measured at 5 K under a magnetic field of 4 T. The inner figure presents the zoom-in data of $R_{\textrm{nl}}^{2\omega}$, for the applied current from 100 $\upmu$A to 300 $\upmu$A.}
		\label{fig:2}
	\end{figure}
	
	
	\begin{figure*}
		\includegraphics[width=440pt]{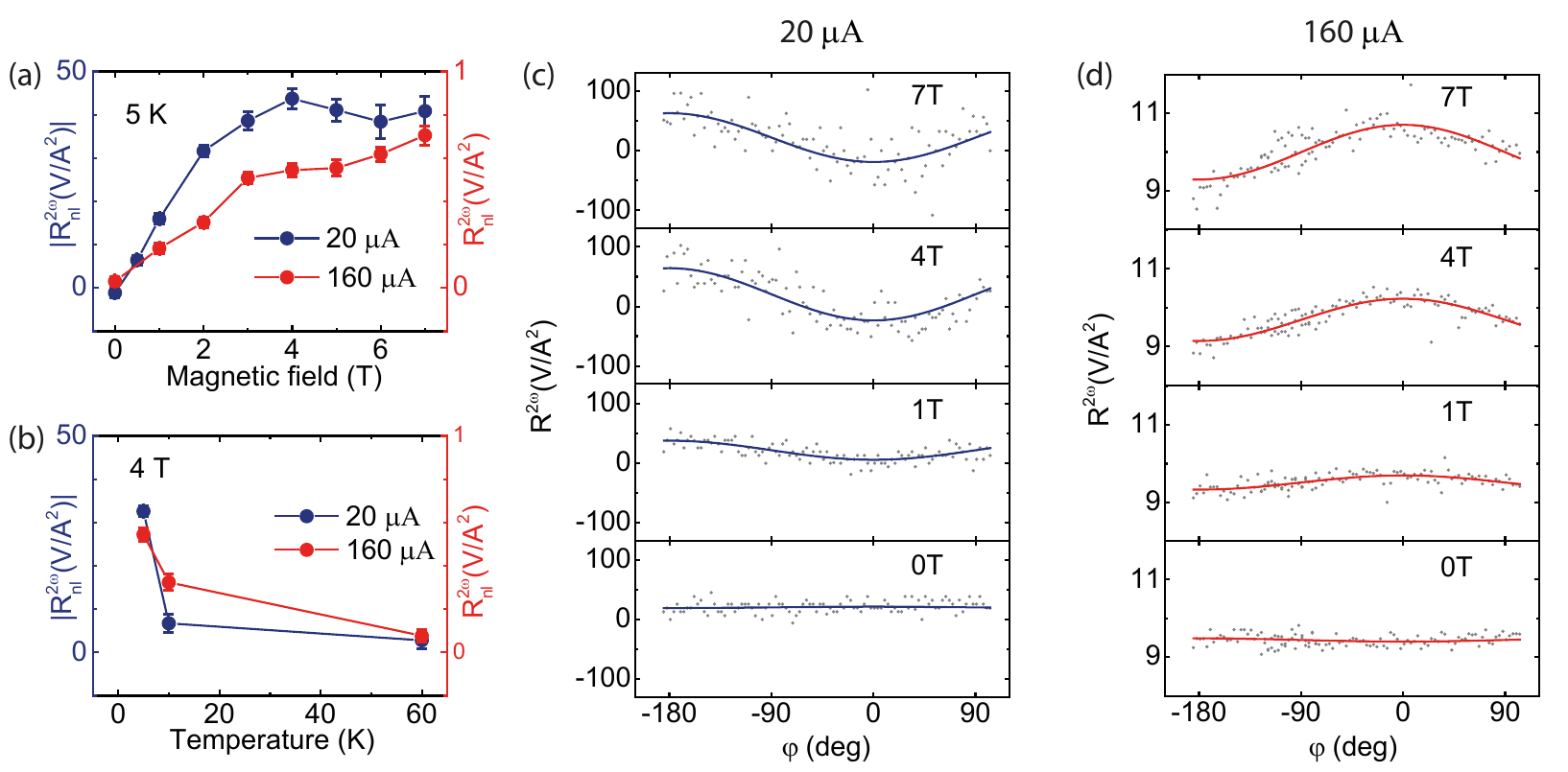}
		\caption{(a). Magnetic field dependence of $R_{\textrm{nl}}^{2\omega}$ with both low current (20 $\upmu$A) and high current (160 $\upmu$A). The fitted cosine amplitude increases with magnetic field until 3 T in both cases. Examples of measured signals are shown in (c) for low bias and in (d) for high bias, with the fitted cosine curves in solid line. (b). The low bias and the high bias signals measured at three different temperatures: 5 K, 10 K and 60 K. The thermal spin signal measured at 10 K is smaller than 5 K for both low bias and high bias cases. }
		\label{fig:3}
	\end{figure*}
\begin{figure*}
		\includegraphics[width=\textwidth]{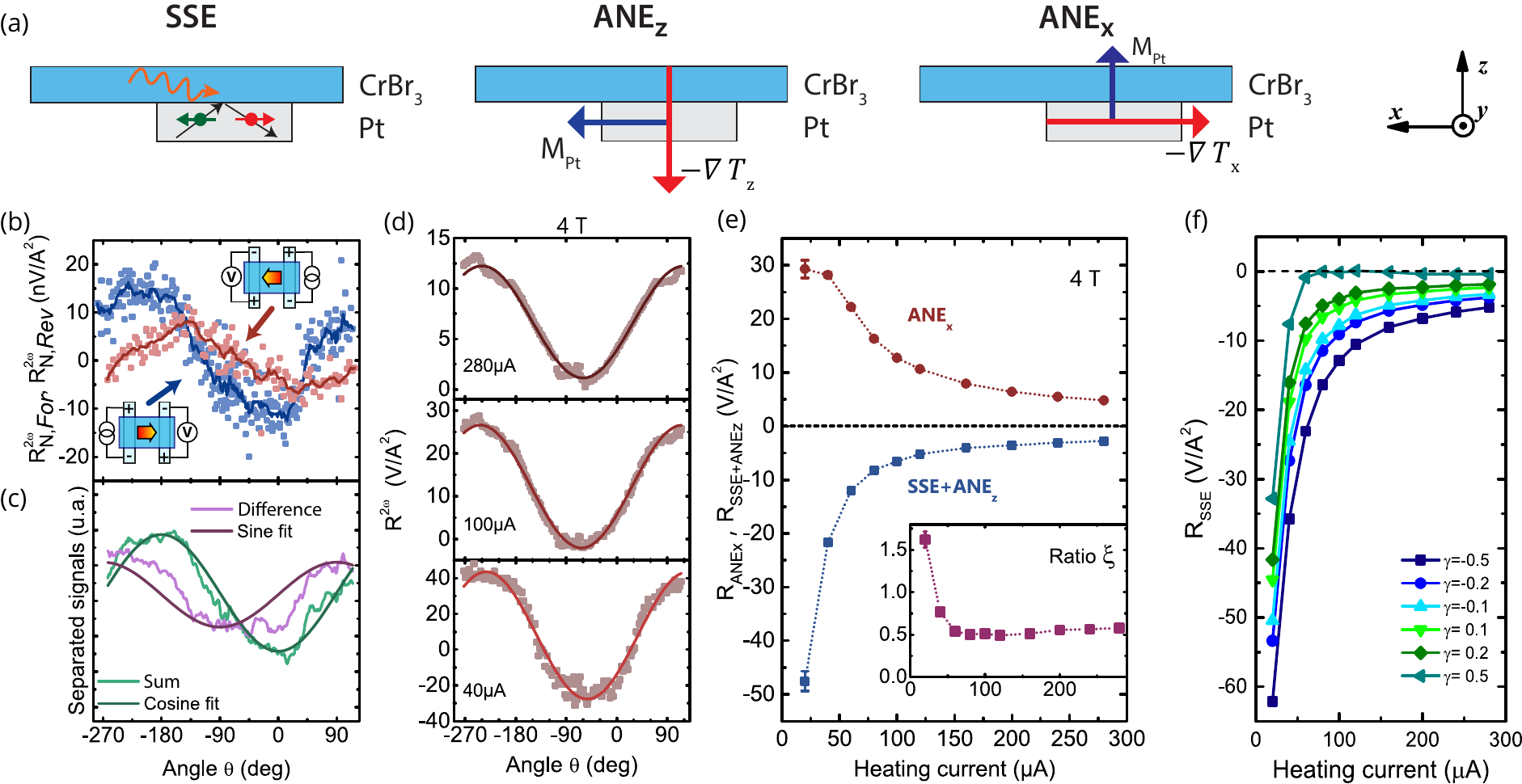}
		\caption{(a) Schematics of the main effects contributing to the detected signal in OOP-nlADMR, $\varphi=0^\circ, \theta \in \left[-180^\circ, 180^\circ\right]$. (b) Second harmonic nlADMR for the forward (blue) and the reverse (red) configurations measured with appliend current of 20 $\upmu$A. (c) Sum $\left(R_{\textrm{NL,For}}^{2\omega}+R_{\textrm{NL,Rev}}^{2\omega}\right)/2$ (green) and difference $\left(R_{\textrm{NL,For}}^{2\omega}-R_{\textrm{NL,Rev}}^{2\omega}\right)/2$ (purple) of the traces in (b), highlighting contributions that are fitted with $cos(\theta)$ and $sin(\theta)$ functions respectively.
			(d) Second harmonic nlADMR shown for 40, 100 and 280 $\upmu$A for an external magnetic field of 4 T rotating in the $x$-$z$ plane. (e) The current dependence of pANE$_x$ (red) and SSE + pANE$_z$ signals (blue) for the forward configuration. In insets of (e) are given the ratio $\xi=-(R_{\textrm{SSE}}+R_{\textrm{pANE}z})/R_{\textrm{pANE}x}$ (bottom inset). (f) Current dependence of the calculated SSE resistance for a range of $\gamma=\partial_z$T / $\partial_x$T.}
		\label{fig:4}
	\end{figure*}

	First and second harmonic responses of the non-local signal have been measured simultaneously all along this study. The first order angular dependence is expected to obey the relation $R^{1\omega} = V/I = R_{0}^{1\omega} + R_{\textrm{nl}}^{1\omega}cos^2{(\varphi)}$  \cite{cornelissen2015long}, where $R_{0}^{1\omega}$ is an offset resistance, and  $R_{\textrm{nl}}^{1\omega}$ is the magnitude of the first harmonic signal. However, we do not observe the expected $cos^2{(\varphi)}$ modulation in the first harmonic signal, as the fitted first order resistance $R_{\textrm{nl}}^{1\omega}$ is only detected in the order of 0.01 m$\Omega$ which is comparable to the standard deviation. An example of measured first harmonic signal can be found in Fig.1(c). Yet, this value is at least 3 orders smaller than the $R_{\textrm{nl}}^{1\omega}$ reported for the Pt/YIG system \cite{cornelissen2015long}). The measurements are carried out over a wide range of applied currents and magnetic fields, and with the maximum lock-in detection sensitivity. A typical measurement of first harmonic non-local signal is shown in Fig.\ref{fig:1}c, for a current of 20 $\upmu$A at 5 K. In contrast, the non-local second harmonic signals exhibit a clear sinusoidal behavior (Fig.\ref{fig:1}e) under an in-plane rotating magnetic field. The magnitudes of non-local signals were fitted with:
	\begin{equation}
	\label{eqn:2}
	R^{2\omega} = \frac{V}{I^2} = R_{0}^{2\omega} +  R_{\textrm{nl}}^{2\omega}cos{(\varphi)},
	\end{equation} where $R_{0}^{2\omega}$ is the offset resistance for the second harmonic signal. A non-zero offset $R_{0}^{2\omega}$ is always present, possibly from unintended Seebeck contribution in the detector\cite{sierra2018thermoelectric}. $R_{\textrm{nl}}^{2\omega}$ is the magnitude of the second harmonic signal. For the corresponding second harmonic measured in Fig.\ref{fig:1}e, we extract an amplitude $R_{\textrm{nl}}^{2\omega}=-36\ \pm \ 1 \ V/A^2$, which is comparable to the magnitude of room-temperature non-local SSE measured on bulk Pt/YIG samples \cite{cornelissen2015long} with equal angular dependence.  
	If we compare to the typical top contact geometry used to detect SSE from YIG \cite{cornelissen2015long}, the same SSE detected here in bottom contact geometry should produce a spin current in the opposite direction. Therefore, the ISHE induced in Pt is reversed compared to the top Pt on YIG, hence, we expect an opposite sign of the signal. The negative sign observed here would correspond to the positive sign measured in \cite{cornelissen2015long} and, if attributed to SSE, reveals a transfer of magnon-spin from CrBr$_3$ to the Pt top surface. However, at this point, we cannot rule out other effects like proximity induced Anomalous Nernst Effect (pANE) in Pt \cite{KikkawaPRL2013}. We discuss about relevant effects later (see Fig\ref{fig:4}c, rotation of out-of-plane magnetic field).
	

	
	

	The current dependence of $R_{\textrm{nl}}^{2\omega}$ is plotted in Fig.\ref{fig:2}, for a contact pair with distance of 950 nm center to center (edge to edge distance of 500 nm). $R_{\textrm{nl}}^{2\omega}$ depends on the applied current non-linearly, and a sign reversal of $R_{\textrm{nl}}^{2\omega}$ occurs between 40 and 100 $\upmu$A. For data measured at 60 $\upmu$A and 80 $\upmu$A, an angular modulation of the second harmonic signal is still observed but it is not described by a simple cosine function \cite{SuppMat}. An example of the negative $R_{\textrm{nl}}^{2\omega}$ at low current is shown in Fig.\ref{fig:2}a (top panel), and an example of the positive $R_{\textrm{nl}}^{2\omega}$ at high current is plotted in Fig.\ref{fig:2}a (bottom panel). The absolute amplitude $\left|R_{\textrm{nl}}^{2\omega}\right|$ in general decreases with increasing current at the heater, as plotted in Fig.\ref{fig:2}b. Its value for positive amplitude at high current is one to two orders of magnitude lower than its value for negative amplitude at low current, depending on the applied current. 
	
	To get better insight of the role of the complex temperature distribution in our device for this non-linear behavior, we employ a 2-dimensional finite element model (FEM) simulating a geometry of the $x$-$z$ plane. Indeed the full hBN-encapsulation of the CrBr$_3$ flake in this device brings inevitable additional heat conduction paths resulting in strong current-dependent thermal gradients in both x and z directions ($\partial_x T$ and  $\partial_z T$ respectively).
	As $\kappa_{\textrm{CrBr}_3}$, the thermal conductivity of CrBr$_3$, is unknown, we ran the computation for different thermal conductance ratios $\eta_{\textrm{K}}$ so that  $\kappa_{\textrm{CrBr}_3}\left(T\right)=\eta_{\textrm{K}} \,\kappa_{\textrm{hBN}}\left(T\right)$, with $\kappa_{\textrm{hBN}}$ the thermal conduction of hBN, and taking into account the highly temperature dependent thermal conduction of the materials (see Supplemental Material VII \cite{SuppMat}). This modeling reveals a strong dependence of the temperature profile as a function of the heating current. It qualitatively supports that the main contribution of the thermal gradient in the Pt detector is in $x$ direction ($\partial_x T$). Yet there also is a non-negligible thermal gradient in $z$ direction ($\partial_z T$), in the CrBr$_3$ as well as in the Pt detector, allowing for SSE and possible unintended effects occurring in the Pt detector that will be discussed below.
	
	%
	
	The in-plane magnetic field dependence on the second order nlADMR amplitude $R_{\textrm{nl}}^{2\omega}$ is plotted in Fig.\ref{fig:3}a. We apply a range of fields from 0 T to 7 T for the in-plane rotation measurements at 5 K. At low current (20 $\upmu$A), we observe a linear increase of $|R_{\textrm{nl}}^{2\omega}|$ from 0 T to 3 T. After 4 T, the magnitude tends to saturate showing only a slight decay (Fig.\ref{fig:3}a and Fig.\ref{fig:3}c). At high current (160 $\upmu$A), we also observe a linear increase of $R_{\textrm{nl}}^{2\omega}$ from 0 T to 4 T, but with magnitudes about 50 times smaller than $|R_{\textrm{nl}}^{2\omega}|$ for low current. After 4 T, the magnitude still increases but at a lower rate (Fig.\ref{fig:3}a and Fig.\ref{fig:3}d). The lower magnitude at high current is consistent with the reduction of the magnetization expected for a temperature increase due to Joule heating.
	The origin of the magnetic field dependence remains unclear. As the saturation of the magnetization of tri-layer CrBr$_3$ in its hard-plane is reported to occur at 400 mT \cite{Kim2019evolution}, the linear increases cannot be simply explained by the saturation of the magnetization as from an isolated CrBr$_3$ layer and reveals the contribution of additional field dependent effects.

	The second order nlADMR is also measured at three different temperatures, 5 K, 10 K and 60 K, and the fitted amplitudes of $R_{\textrm{nl}}^{2\omega}$ are shown in Fig.\ref{fig:3}b for low (20 $\upmu$A) and high current (160 $\upmu$A) measured under 4 T. Compared with the signal at 5K, the fitted cosine amplitude at 10K decreases for both low and high bias. Far above T$_c$ at 60 K, a very small but non-zero value of $R_{\textrm{nl}}^{2\omega}$ is observed in our measurements (0.08 $\pm$ \ 0.03 \ $V/A^2$ at 160 $\upmu$A and -3 $\pm$ \ 2\ $V/A^2$ at 20  $\upmu$A). We attribute this small non-zero value to an artifact from the measurement setup \cite{SuppMat}.
	
	%
	We present hereafter a series of out-of-plane nlADMR (OOP-nlADMR) measurements i.e. fixing $\varphi=0^{\circ}$ and varying $\theta$ by rotating the magnetic field in the $x$-$z$ plane, as defined on Fig.\ref{fig:1}. Some examples and the current dependence of this OOP-nlADMR are summarized in Fig.\ref{fig:4}. 
	The first observation, with Fig.\ref{fig:4}b and \ref{fig:4}d as examples, is that 
	all OOP-nlADMR signals exhibit a non-zero angular phase shift varying with the heating current. We investigated the origin of this phase considering 
	the various effects that could add to the SSE signal. Nernst, Seebeck, Spin Nernst Magnetoresistance (SNMR)\cite{Meyer2017,Kim2017} effects are discarded as major contributions, either due to the probing geometry, or their angular dependence, a detailed description is given in \cite{SuppMat}. However the anomalous Nernst effect (ANE), which has already been reported as a possible effect, arising from a proximity induced ferromagnetism into the Pt  \cite{Leutenantsmeyer2016,Zollner2016,Meier2015,Guo2014,Lu2013,KikkawaPRL2013}, cannot be ruled out.

	Considering a proximity ANE (pANE) in Pt, a transverse pANE voltage $\Delta V_{\textrm{pANE}}$ reads :
	\begin{equation}
	\frac{\Delta V_{\textrm{pANE}}}{L_{\textrm{Pt}}}=\left |\mathbf{\nabla} V \right |_y = \left |-S_{\textrm{pANE}}\left(\mathbf{m}\times \left[-\mathbf{\nabla} T\right]\right)\right |_y
	\label{eqn:ANE}
	\end{equation}
	Where $S_{\textrm{pANE}}$ is the pANE coefficient, $\mathbf{m}$ is the unit vector of direction of the magnetization and $L_{\textrm{Pt}}$ is the $y$-axis length of the contact area of Pt with CrBr$_3$. As the magnetization of CrBr$_3$ is expected to saturate for fields beyond 1 T in the hard plane \cite{Richter2018,Kim2019evolution}, we also assume the proximity induced magnetization parallel to the magnetic field at 4 T. Then, two contributions of the pANE are distinguished (Fig.\ref{fig:4}a) : the pANE signal caused by the IP gradient $\partial_x T$, pANE$_x$ which varies as $sin\left( \theta \right)$, and the pANE signal caused by the OOP gradient $\partial_z T$, pANE$_z$ which varies as $cos\left(\theta\right)$.

	The pANE induced by the temperature gradient along $x$ (pANE$_x$) can be isolated from the other signals 
	by changing the heat flow direction. By interchanging the heater and detector contacts, the heat flow direction along the $x$ axis ($\propto\partial_x T$) is reversed, but the heat flow direction along the $z$ axis ($\propto\partial_z T$) remains the same. Hence, the pANE$_z$ contribution will stay unchanged while the pANE$_x$ will reverse its sign. In Fig.\ref{fig:4}b, we provide a normalized second order nlADMR $R^{2\omega}_N=R^{2\omega} A_{\textrm{Pt}}/L_{\textrm{Pt}}$, with $A_{\textrm{Pt}}$ the Pt electrode cross-section, at 20 $\upmu$A and 4 T, for the configuration forward defined in Fig.\ref{fig:1}, and the nlADMR from a reversed geometry where heater and detector are interchanged. As the width and length of the two electrodes are different, as well as their interface with CrBr$_3$ possibly, the heating power injected will differ by a small factor. Therefore our comparison remains only qualitative. Nevertheless, the amplitudes and offsets are alike and the two traces differ mainly by the apparent opposite phase shift.
	
	If both pANE$_x$ and pANE$_z$ contributions are significant in our system, the difference between the forward geometry (Fig.\ref{fig:4}b) signal and the reverse geometry (Fig.\ref{fig:4}b) signal will reveal the $sin\left(\theta\right)$ behavior, and the sum of these two signals will reveal the $cos\left(\theta\right)$ behavior. As a result, we obtain the respectives traces shown in Fig.\ref{fig:4}c. The good agreement of the fittings on both curves is a confirmation that the pANE is present in the Pt detector.
	
	Based on this observation, we extracted the two contributions for every ADMR at different current and at a constant magnetic field of 4 T, by fitting the expression $R^{2\omega}=R_{0}^{2\omega}+R_{\textrm{SSE}+\textrm{pANE}z}\, cos \theta +R_{\textrm{pANE}x}\, sin \theta$. The measurements at 40, 100 and 280 $\upmu$A  are shown in Fig.\ref{fig:4}c, and the fitted sinusoidal curve presents the phase shift in each case. The current dependence of the extracted amplitudes is provided in Fig.\ref{fig:4}d. The $R_{\textrm{SSE}+\textrm{pANE}z}$ and $R_{\textrm{pANE}x}$ contributions both follow a similar decreasing trend with applied current. While $R_{\textrm{SSE}+\textrm{pANE}z}$ dominates at 20 and 40 $\upmu$A, $R_{\textrm{pANE}x}$ becomes close to twice $R_{\textrm{SSE}+\textrm{pANE}z}$ at higher current. The variation of the amplitude of $R_{\textrm{SSE}+\textrm{pANE}z}$ at low currents follows the variation of the signal for IP field rotation in Fig \ref{fig:2}b, however the sign reversal for the derived $R_{\textrm{SSE}+\textrm{pANE}z}$ does not occur in the OOP configuration.
	
\begin{figure}
		\includegraphics[width=240pt]{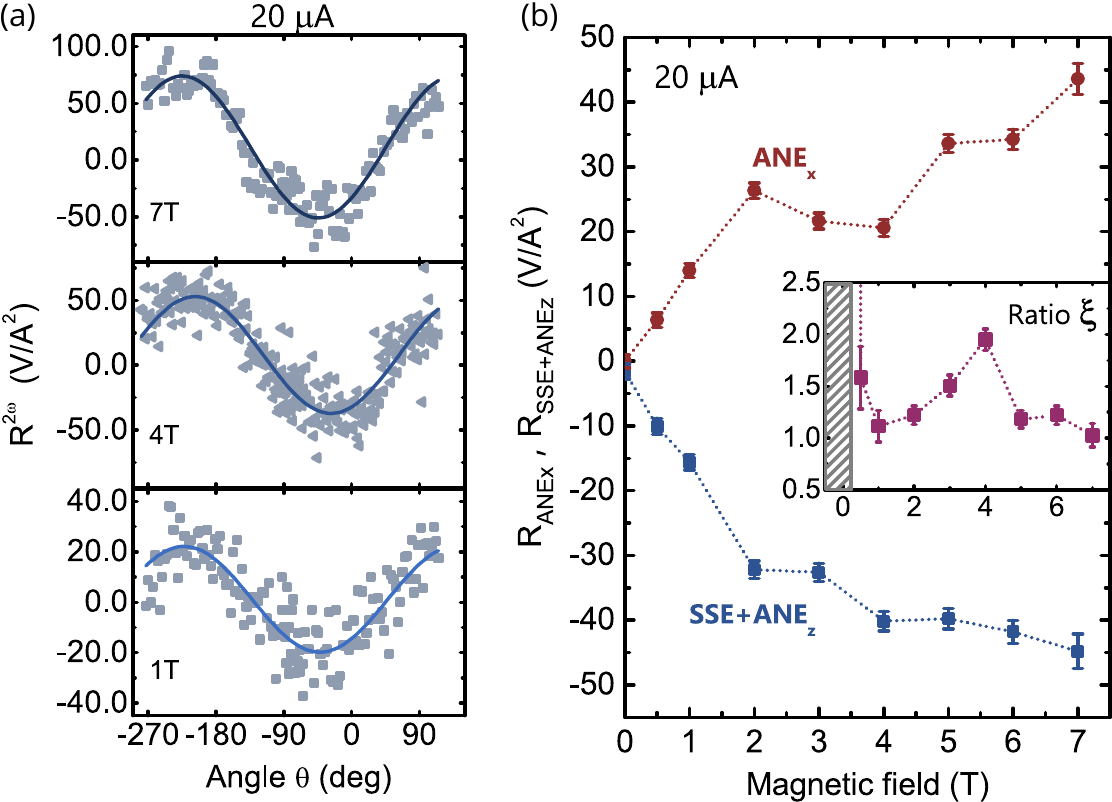}
		\caption{(a) SSE angular dependence shown for 1, 4 and 7 T, with current fixed to 20 $\upmu$A at 5K. (b) Magnetic field dependence of pANE$_x$ and SSE + pANE$_z$ signal amplitude for the forward configuration. In the inset of (b) the ratio $\xi = -(R_{\textrm{SSE}}+R_{\textrm{pANE}z})/R_{\textrm{pANE}x}$ is given. See \cite{SuppMat} for the data extraction in detail.}
		\label{fig:5}
	\end{figure}

	To elucidate the contribution of the spin Seebeck, we introduce the ratio $\xi=-R_{\textrm{SSE}+\textrm{pANE}z}/R_{\textrm{pANE}x}=-\left(R_{\textrm{SSE}}+R_{\textrm{pANE}z}\right)/R_{\textrm{pANE}x}$
	of the two contributions (inset of Fig.\ref{fig:4}e), the ratio $\delta = S_{\textrm{pANE}}^z$/$S_{\textrm{pANE}}^x$ to account for any difference between the IP ($S_{\textrm{pANE}}^x$) and OOP ($S_{\textrm{pANE}}^z$) proximity anomalous Nernst coefficients, as well as the ratio $\gamma=\partial_z T/\partial_x T$ of the temperature gradients in Pt. As a result, the SSE contribution to the measured signal simply reads (demonstration in \cite{SuppMat}):
	\begin{equation}
	R_{\textrm{SSE}}=R_{\textrm{pANE}x} \left(\delta \gamma - \xi \right)
	\label{equ:evalSSE}
	\end{equation}
	Based upon the fact that the saturated magnetization of CrBr$_3$ has been reported to be of same magnitude when oriented IP or OOP, we assume $\delta\approx 1$, i.e. $S_{\textrm{pANE}}^z \approx S_{\textrm{pANE}}^x$. Following this assumption, the estimated ratio of the two contributions $\gamma$ lays between $-0.20$ and $0.15$, according to our FEM simulation based on thermal conduction properties of CrBr$_3$ and hBN layers (i.e. the ratio $\eta_{\textrm{K}}=\kappa_{\textrm{CrBr}_3}/\kappa_{\textrm{hBN}}$ ) (see details in \cite{SuppMat}). Even using $\delta \gamma = \pm0.5$ accounting for the possible underestimation of $\partial_z T$ due to the omission of a small heat leakage via the Pt/Au contacts leads on SiO$_2$, we extract a significant SSE contribution to the nlADMR signal at low heating current, as plotted in Fig.\ref{fig:4}f.
	%
	We provide the magnetic field dependence of the OOP-nlADMR in Fig.\ref{fig:5}. Fig.\ref{fig:5}a shows examples of the evolution of the OOP-nlADMR for 1, 4, 7 T, for a  current fixed to 20 $\upmu$A. The same operation to separate pANE$_z$+SSE from pANE$_x$ is applied to this measurement set and the amplitude variation of each component is shown in Fig.\ref{fig:5}b for magnetic fields from 0 to 7 T. The pANE$_z$+SSE variation is comparable to the one measured in in-plane rotation configuration (Fig.\ref{fig:3}a, blue curve), except that we do not observe the high field saturation decrease. The dependence of the pANE$_x$ trace follows a similar increase until 2 T, but shows a slight decrease for 3 and 4 T and increases again to reach the same value as pANE$_z$+SSE at 7T. This behaviour is captured into the $\xi$ ratio that shows a peak above 1.5 for 3 and 4 T and a value remaining around 1 for other fields strengths. As the temperature profile is fixed, the difference between SSE+pANE$_z$ and pANE$_x$ must be strongly linked to the magnetic properties of the CrBr$_3$/Pt structure. 
	\section{Discussion}
	By analyzing the OOP-nlADMR, we show that pANE$_x$ presents a different angular dependence than SSE and pANE$_z$ allowing to separate the two contributions. Despite the lack of insight on the mechanism inducing the magnetization in Pt, this assumption is based on that the magnetic moments emerging on the Pt atoms are imprinted by the moments of CrBr$_3$. Yet the saturated magnetization of CrBr$_3$ has been measured to differ by less than 20\% between the orientation along the easy axis and the orientation in the hard plane. Therefore, the induced magnetization in Pt is expected to behave accordingly, leading to a comparable anomalous Nernst coefficient depending on the magnetization value but weakly on its orientation.
	
	A pANE contribution to the ADMR has been identified in Pt/YIG systems as well, but the pANE$_z$ represents at most 5\% of the voltage signal, the left 95\% being attributed to SSE induced ISHE \cite{KikkawaPRL2013}. Because of the significant magnetic exchange field already noticed in CrBr$_3$ \cite{Tang2019,Kim2019a} as well as the strong temperature gradients involved (beyond 2 orders of magnitude higher than in \cite{KikkawaPRL2013}), in our CrBr$_3$/Pt system, the pANE cannot be neglected and the SSE signal is at best comparable with the pANE$_z$.

	In Pt/YIG system, the magnon SSE signal decreases with the magnetic field  \cite{cornelissen2016field}. In Fig\ref{fig:3}.a, we notice that, after 3T, the fitted amplitude of the low current curve does not change with magnetic field, but $R_{\textrm{nl}}^{2\omega}$ of the high current curve increases linearly with magnetic field. In other words, $R_{\textrm{nl}}^{2\omega}$ at low current tends to decrease where SSE contributes most, compared to the amplitude at high current where the SSE contributes less. Hence, our measurements, with support of a temperature distribution simulation, suggest that the high amplitude signal observed at low current is dominated by SSE from CrBr$_3$.

	According to the expected angular dependence of the SSE and ANE, the SSE+pANE$_z$ signal should appear in both IP-nlADMR and OOP-nlADMR while pANE$_x$ should be only detected in OOP-nlADMR. Therefore, the same current dependence of SSE+pANE$_z$ in both configurations is expected. According to the FEM simulation, a reversal of $\partial_z T$ occurs at sufficiently high current, simultaneously in CrBr$_3$ and Pt at the detection interface (see \cite{SuppMat}). This leads to a reversal of the SSE+pANE$_z$, most likely dominated by ANE in the high current range. However, the sign reversal is only observed in the IP measurements (in Fig\ref{fig:2}b), not showing in the OOP measurements after the separation (in Fig\ref{fig:4}e). As the IP and OOP measurements were performed with different cool-down processes, the thermal conductivity is possibly changed at the interface. This implies that the sign reversal current is possibly shifted to a much higher value, therefore not observed in the OOP measurements. 
	
	Furthermore, we also suggest that a quantitative discrimination between pANE and SSE is possible. We provide an indicative estimation of the magnitude of the SSE based on the assumption that the pANE coefficient is equal for ANE$_x$ and ANE$_z$. By far, we are limited by the current knowledge on the material properties of the 2D magnet. However, if the thermal conduction profiles and the magnetization dynamics are characterized concretely, a more accurate separation of the two spin-caloritronic effects can be realized.
	
	Nevertheless, the magnetic field dependence of pANE$_x$ and SSE+pANE$_z$ and the difference between them bring new questions. The ANE scales with the magnetization via the coefficient $S_{\textrm{pANE}}$. The non-monotonic field dependence of pANE$_x$ suggests a complex evolution of the induced magnetization in Pt, due to either the presence of magnetic domains or any additional interaction at the interface.
	
	\section{Conclusions}
	To conclude, we demonstrate the relevance of the full hBN encapsulation and the bottom contacting design to enable the integration of air-reactive materials such as CrBr$_3$, for studying spin-caloritronic effects in 2D magnets. By using second order nlADMR measurement on such an encapsulated CrBr$_3$/Pt device, we reveal, by detecting the presence of a proximity ANE voltage, a significant proximity induced magnetism from CrBr$_3$ into the adjacent Pt contacts. With reasonable assumptions, we conjecture about the presence of a weak SSE, dominating the signal in the low current regime, while the pANE prevails for currents above 60$\upmu $A. The non-trivial magnetic field dependence of the separated effects leaves open questions as for the current understanding of magnetic effects at the interface of heavy metal and 2D magnets. The encapsulation shows itself an elegant technique to address these questions in deeper investigations of air-degradable 2D materials, and opens the way to future magnon transport studies.

	\section{Methods}
	CrBr$_3$ and hBN crystals are provided by a commercial company HQgraphene. CrBr$_3$ is an air sensitive material. To study magnonics with CrBr$_3$ in a non-local geometry, we encapsulate a 7 nm-thin chromium tribromide flake and platinum (Pt) strips into two hexagonal Boron Nitride (hBN) layers (top layer and bottom layer). The stacking of van der Waals materials was performed in a glove box filled with inert gas argon by using standard PC/PDMS dry transfer method. Pt strips were first grown on bottom hBN. After that CrBr$_3$ with a top hBN thin layer was transferred on top of the Pt strips. See \cite{SuppMat} for more details in fabrication process.

	\begin{acknowledgments}
		The authors thank Prof. Justin Ye and Puhua Wan for granting us access to their transfer system in an Ar glove box. We thank Johan G. Holstein, Hans de Vries, Herman Adema, and Tom J. Schouten for technical assistance. We thank fruitful discussions with Geert R. Hoogeboom, Alexey A. Kaverzin and Jing Liu. This project has received funding from the Dutch Foundation for Fundamental Research on Matter (FOM, now known as NWO-I) as a part of the Netherlands Organization for Scientific Research (NWO), the European Union’s Horizon 2020 research and innovation programme under grant agreement No 696656 and 785219 (Graphene Flagship Core 1 and Core 2) and Zernike Institute for Advanced Materials. MHDG acknowledges support from NWO VENI 15093.
	\end{acknowledgments}

\end{document}